\documentclass{ieeeaccess}
\usepackage{cite}
\usepackage{amsmath,amssymb,amsfonts}
\usepackage{algorithmic}
\usepackage{graphicx}
\usepackage{textcomp}
\def\BibTeX{{\rm B\kern-.05em{\sc i\kern-.025em b}\kern-.08em
    T\kern-.1667em\lower.7ex\hbox{E}\kern-.125emX}}

\usepackage{subfigure}
\usepackage{pdflscape}
\usepackage{amsmath}

\usepackage{lineno,hyperref}

\usepackage{caption,setspace}
\begin{document}
\history{Date of publication xxxx 00, 0000, date of current version xxxx 00, 0000.}
\doi{10.1109/ACCESS.2017.DOI}

\title{Analysing Emergent Users' Text Messages Data and	Exploring its Benefits}

\author{\uppercase{Anas Bilal}\authorrefmark{1},
\uppercase{Aimal Rextin\authorrefmark{1},		
\uppercase{Ahmad Kakakhail\authorrefmark{1}
and Mehwish Nasim }\authorrefmark{2}}}

\address[1]{Department of Computer Science, COMSATS University Islamabad, Islamabad, Pakistan }
\address[2]{ARC Centre of Excellence for Mathematical and Statistical Frontiers, University of Adelaide, Australia}


\markboth
{Bilal \headeretal: Analysing Emergent Users' Text Messages Data and	Exploring its Benefits}
{Bilal \headeretal: Analysing Emergent Users' Text Messages Data and	Exploring its Benefits}

\corresp{Corresponding author: Aimal Rextin (e-mail: aimal.rextin@comsats.edu.pk).}

\begin{abstract}
While users in the developed world can choose to adopt the technology that suits their needs, the \textit{emergent} users cannot afford this luxury, hence, they adapt themselves to the technology that is readily available. When technology is designed, such as the mobile-phone technology, it is an implicit assumption that it would be adopted by the emergent users in due course. However, such user groups have different needs, and they follow different usage patterns as compared to users from the developed world. In this work, we target an emergent user base, i.e., users from a university in Pakistan, and analyse their texting behaviour on mobile phones. We see interesting results such as, the long-term linguistic adaptation of users in the absence of reasonable Urdu keyboards, the overt preference for communicating in Roman Urdu and the social forces related to textual interaction. We also present two case studies on how a single dataset can effectively help understand emergent users, improve usability of some tasks, and also help users perform previously difficult tasks with ease. 
\end{abstract}


\titlepgskip=-15pt

\maketitle

\section{Introduction}

\label{sec:intro}

Low cost mobile phones have allowed the wide scale adoption of smart phones in developing countries; for example, South Asian users alone  constitute one of  the largest mobile phone user bases in the  world. In the last few years, researchers, have referred to user groups from developing countries as \textit{emergent users}. Such user groups are either less educated, economically disadvantaged, geographically dispersed, or have a culturally heterogeneous background \cite{joshi2013technology}. Millions of people in the developing world own mobile phones and large proportions of users have access to smartphones, owing to the availability of low cost smartphones. Technology is designed primarily for  the needs of a user from the developed world and it is assumed  emergent users will  adopt it in a similar fashion. However, studies show that emergent users adopt these technologies in different ways \cite{joshi2013technology} \cite{Nasim17}, with their unique usage peculiarities \cite{nasim2016understanding, pearson2017evaluating}.   Jones et al. \cite{jones2017beyond} argued that  emergent users should be engaged in technology design to yield designs of greater value.

Studies  recommend that the interface design follows the traditional life cycle of HCI. However, this approach needs both time and financial resources; both lacking in case of emergent users. We propose an alternative approach; i.e., data driven usability improvement. This approach is feasible in today's world due to the large amount of digital footprints left by users on their computing devices. We present two case studies on how a single dataset can help understand emergent users and improve  usability of some tasks.   

We decided to use data from text messages for our analysis because use of text messaging  has increased in recent years. Pakistan Telecommunication Authority \cite{pta} estimates that there were $133.24$ million mobile subscribers in $2016$ and  more than $301.7$ billion text messages were sent in $2014$. We observed that Pakistani users generally use Roman Urdu when communicating by text messages.  \emph{Roman Urdu} is the colloquial term given to the practice of writing Urdu (the national language of Pakistan) in Roman script. It is generally agreed by local users that this practice evolved when users who were not comfortable in English started  to communicate over text messages \cite{ref4,ref2,ref13}.

\subsection{Contributions} \label{sec:contributions}


In this paper, we extend the work that we presented earlier \cite{bilal2017roman} by performing additional analysis on the collected dataset. This dataset\footnote{Our dataset is available for academic use at  \url{https://github.com/CIIT-HCI/Roman-Urdu-SMS-Corpus}}  is a collection of original text-messages corpus of $116$ mobile phone users in Pakistan. This data set consists of a hand-crafted file that includes grouping of various spelling variations of the same word. The following are our contributions for this extended study:

\begin{enumerate}

	\item \textbf{Texting behaviour:} Our first contribution is studying the texting behaviour of Pakistani users, such as, the predominant language used by the users and the various spelling variations of different words. This will be covered in Section \ref{sec:trends}.
	\item \textbf{Spelling Variations: } Since Roman Urdu has naturally evolved, there are no standardized spellings in place. In Section \ref{sec:spelling_variations}, we will show that spelling variations have three main categories.
	\item \textbf{Case Studies:} The marked contribution of this paper is to show the importance of collecting datasets from developing countries in order to understand the use of technology in understudied populations. In this regard, we will discuss the following case studies:
	
	\begin{enumerate}
		\item Availability of word completion feature is a major ease for smartphone users. In Section \ref{sec:word_completion}, we show that by using this corpus one can help emergent smartphone users by providing him/her with more accurate word completion.	
		\item In Section \ref{sec:intimate}, we will discuss how users' text data can be used to understand  the social dynamics governing the interactions among people in developing countries, especially the ones involving intimate relations.
		
	\end{enumerate}
\end{enumerate}

\subsection{Background} \label{sec:related_work}

Users of computing systems who are disadvantaged are generally referred to as \emph{emergent users} \cite{joshi2013technology}. These include users from developing countries who face many challenges due to their different cultures and language, low education, and late access to technology etc. The number of mobile phone users from developing countries have increased dramatically in recent years due to the decreasing cost of smartphones and mobile internet. This has led to  a steady stream of studies that analyse problems and potential solutions for these users.  These include studying text messages from a structural and functional point of view \cite{bilal2017roman};   a study exploring  the use of computing technology by  young migrant workers in China \cite{lang2010social}; and  a job searching website designed for illiterate people in Pakistan \cite{khan2017job}. 

It has been argued that emergent users especially illiterate users find it difficult to use text based features on phones \cite{thies2015user}. Various studies suggest that voice based access of such features should be made easier, e.g., voice-based use of text messages  \cite{friscira2012getting} and a  voice-based job searching mobile application for Pakistani illiterate users  \cite{raza2013job}. However, text based communication has its own advantages such as its asynchronous nature and privacy, making it  more convenient under certain social settings. The benefits of text messaging and their difficulty in using them led Pakistani users to  adapt to the situation  by using   Roman script to convey their messages in the local language; this style of writing is known as \emph{Roman Urdu} \cite{bilal2017roman}. Any kind of analysis on  Roman Urdu is made complex by the fact that Roman Urdu has no standard spelling defined and different people may use different spellings for the same word \cite{ref15, ref13, ref17}. 

Text messaging communication is becoming more popular day by day, hence attracted the attention of researchers for analysis. For example some studies suggest that text messages can be classified into two disjoint groups. The first class of messages are called  \emph{informational messages} and they contain information of a practical nature, and the second class is called  \emph{relational messages} and they contain greetings and personal conversations etc. \cite{thurlow2003generation, ring2005mobile, faulkner2005when}. Some other studies compare the structure and function of the text messages of users with different characteristics such as age, gender, experience etc. \cite{bernicot2012forms,goumi2011smslength, deumert2008mobile, thurlow2003generation,crosswhite2014texting}. There are even studies that look into how the language used on Facebook  or WhatsApp etc. is different from standard languages \cite{verheijen2016} and studies that  examine the functions of emojis in one-to-one messaging via text \cite{cramer2016sender}.

\subsubsection*{Roman Urdu Corpora}


Urdu is not only the national language of Pakistan but also the official language of many Indian states and is among the most widely spoken languages of the sub-continent \cite{ref5, ref15, ref17}. Researchers have collected various corpora of Roman Urdu in natural settings  including a collection of $82,000$ tweets from Pakistan and using it to design an algorithm to separate English words from Roman Urdu words \cite{ref14}. Hussain collected a corpora of $5000$ text messages and analysed the linguistic patterns of emergent users \cite{ref2}. There is also a corpus of  $50,000$ Roman Urdu SMS and a Roman Urdu corpus containing $1$ Million messages from chat rooms \cite{ref4}. We can see that the larger datasets are from the web chat rooms, which represents a totally different setting than text messages. 

There have also been a few studies that investigate possible applications such as bilingual classification and sentiment analysis \cite{ref6}; transliteration \cite{ref13,kamran2010transliterating}; word prediction \cite{ref7}; and tagging parts of speech etc.

There are several applications of  natural languages corpus  such as detecting  SPAM emails and messages\cite{sohn2012content,liang2015trusms,hidalgo2006content}  and other natural language processing tasks. Hence it is no surprise that various researchers have collected natural language corpora. For example, Tagg \cite{tagg2013language} collected a large scale corpus and performed linguistic investigation on about $11,000$ text messages.

\subsection{Organization}

Section \ref{sec:intro} is the introduction and motivation. Section \ref{sec:data_collection} describes our dataset. In Section \ref{sec:trends}, we discuss the texting behaviour of our user group, whereas in Section \ref{sec:spelling_variations} we analyse spelling variations in Roman Urdu text messages. In Sections \ref{sec:word_completion} and  \ref{sec:intimate}  we report our two case studies. Finally, Section \ref{sec:conclusions} concludes our paper. 

\section{Dataset} \label{sec:data_collection}
There are several ways to textually communicate via a mobile phone including SMS, WhatsApp, Viber, Email etc. In the planning stages of this research work, we wanted to collect text data both from SMS and  WhatsApp. However, due to technical difficulties in  accessing WhatsApp data (mainly API restrictions), we decided to check whether our research objectives can be fulfilled by SMS data. This was done by conducting a pretest. We presented a group of users a single  question: \textit{``How  do they perceive their communication is divided among various channels including WhatsApp, phone calls, SMS, and other less popular channels of communication"}. The  results showed that on average our participants perceived that  $23.6\%$  of their communication was done through calls, $39.6\%$ through WhatsApp, $15.9\%$ through other less popular text communication channels like Facebook Messenger and finally $20.9\%$ through SMS. Hence, SMS accounts for  $27.36\%$ of all text communication channels on smartphones. This gave us the confidence to collect a large enough data set in order to derive useful results.

Initially, we wanted to acquire text messages data from mobile service providers.  However, we faced two constraints in this regard. The first was that  legally, text content is stored for only $7$ days by the local service providers after which the content is deleted. The second constraint was that the service providers were reluctant to share their data   due to various security and privacy concerns\footnote{Knowledge acquired through private communication, 2016.}. Our dataset was collected through a custom \texttt{Android} application specifically developed for this study. {We gathered $346,455$ individual SMS messages from $116$ students of a local university. These $116$ egos exchanged text messages with $2294$ alters \footnote{\textit{Ego} is an individual who installed our application and \textit{alters} are his/her contacts.}. Hence, our results are not only valid for the limited number of participants but also for all their alters with whom they exchanged messages. }We collected statistics such as frequency of alphabets, message length, time of message, unique words etc.\footnote{The data collection process was approved by the Research Ethics Committee at the local university. Details can be provided upon request.}. We did not collect any data that would compromise the privacy of the users such as the complete message contents. We were conscientious  from the start of the experiment about the need to protect personal information of our participants. We took the following measures to ensure that:

\begin{enumerate}
	\item We did not store  any personal information like phone numbers and names and instead stored unique codes to identify the various alters for each ego. 
	\item We did not gather any  significant information about the text messages content rather, we gathered individual words and word bigrams but they were stored alphabetically for each individual's full text messages history. This made it impossible to reconstruct any meaningful information from them. 
	\item Participants were given the autonomy to share their messages with us. A spreadsheet file was generated on their smartphones. They could then analyse it before sending it to us by email. It was clearly communicated to them that they could delete some or all of their data had they felt uncomfortable sharing it. 
\end{enumerate}

We collected SMS data from $53$ females and $63$ males who were students in a local university. Our average participant had $13.41$ outgoing messages and $13.92$ incoming messages in a typical day. There are a total of $446,483$ words including both Roman Urdu and English words. This indicated that SMS is still popular despite the popularity of other messaging applications such as WhatsApp etc.

 We  gathered  subjective information from our participants through  a questionnaire whose  results were partially discussed in our preliminary paper \cite{bilal2017roman}. One of the questions that we asked was \emph{whether they deleted any data before initiating the data collection process}. We found that $46.13\%$ of our participants deleted either $1$ or $2$ complete conversions with particular alters. The mean number of conversations deleted came out to be $0.92$. The participants informed us that these alters were either very close or intimate friends. We will further discuss this aspect in Section \ref{sec:intimate}. 

\section{Initial Analysis}\label{sec:trends}
This section describes some important artefacts of the data specifically revolving around the dominant language in text messages. We studied whether the choice of language varies from alter to alter, and the degree to which it conforms to reciprocity in communication. We then studied the extent of spelling variations of a single word. We  also looked at the different textisms present in text messages. 
\subsection{Choice of language} \label{sec:language_choice}

We asked our participants about their preferred language for textual communication. We found at a $95\%$ confidence level that, $73.2\% \pm 6.58\%$ generally type their messages  in Roman Urdu. The same was confirmed from analysing $30$ uniformly random words from each participant, making a total of $2896$ words. The primary reason driving their preference was ease of conveying their message as well as the perception that the message will be understandable by the recipient.

We also found at a confidence level of $95 \%$ that $82.8\% \pm 6.4\% $ participants  use Roman Urdu when sending a text message to some alters and English for others. This could be attributed to many reasons, specifically to the nature of the ego's relationship with the alter. 
\subsection{Reciprocity}
There is a strong evidence that people naturally tend to match both the alter's vocabulary and sentence structure, when in a dialogue \cite{koulouri2016and}. This natural alignment is called  \emph{reciprocity}. We wish to test whether reciprocity is present in our dataset, i.e. whether some ego-alter pairs predominantly converse in English while others predominantly converse in Roman Urdu?


\begin{figure} [!t]
	\centering
	\includegraphics[width=.9\linewidth] {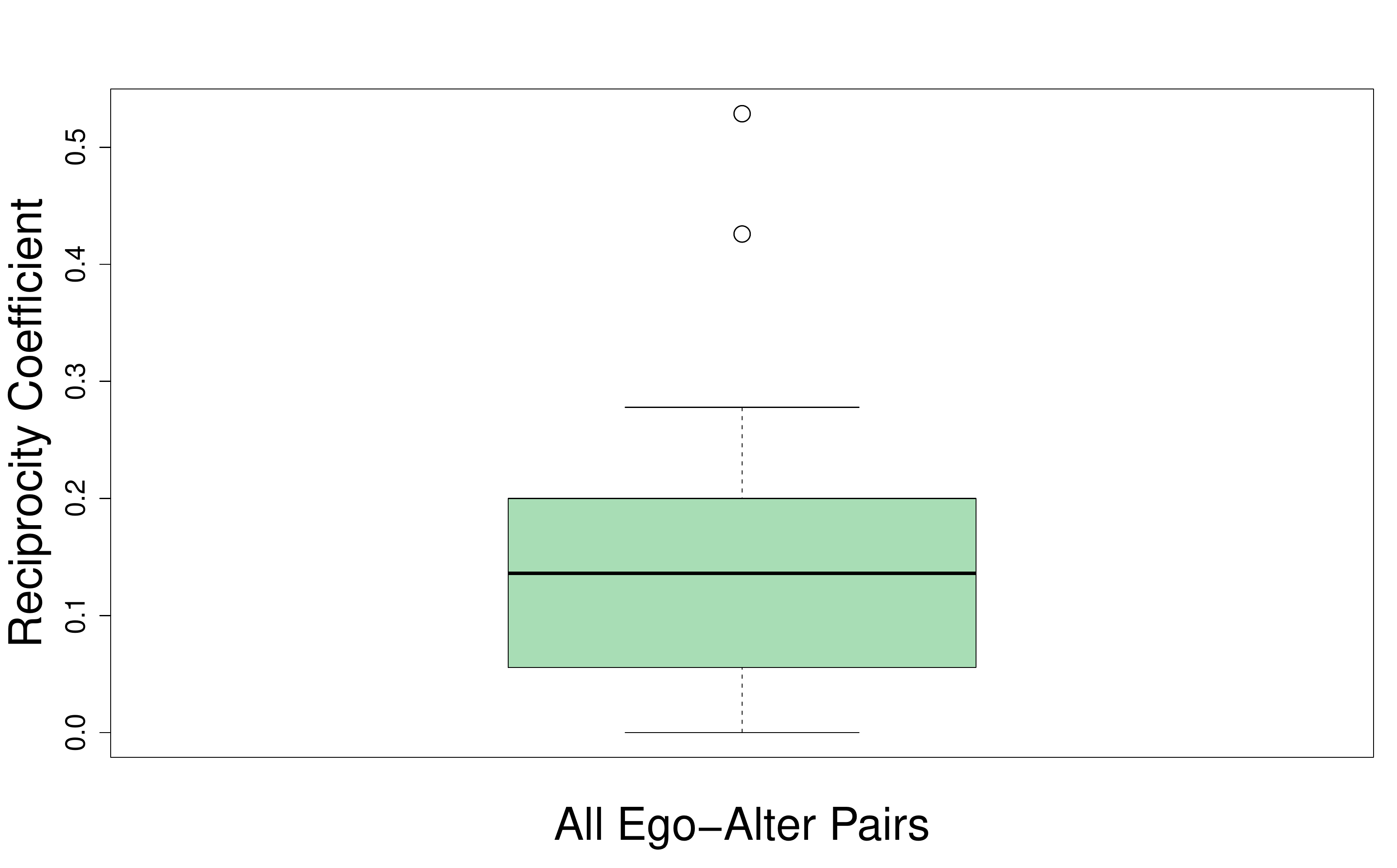}
	\caption{Boxplot showing the distribution of the reciprocity  coefficient  $p$ for all ego-alter pairs. We can see that most ego-alter pairs have low value of $p$, indicating that most words exchanged between an ego-alter pair tend to be in one language or the other.}
	\label{fig:p_distribution}
\end{figure}

We generated a uniform random sample of $96$ ego-alter pairs out of a total of $11942$ such pairs.  We then obtained all unique words that were exchanged between these ego-alter pairs and manually assigned them a language label\footnote{Roman Urdu or English}. Based on this we computed two quantities: $p_s$, the proportion of words sent by the ego to the said alter in English, and $p\textsubscript{r}$, the proportion of words received by the ego from the said alter in English. Based on these we computed the following:
\begin{equation}
p=|p\textsubscript{s}-p\textsubscript{r} |
\end{equation}
We call $p$ as the \emph{reciprocity coefficient} of a particular ego-alter pair \footnote{We will get the same value of $p$ if we  calculate it w.r.t. to Urdu instead of English}. Fig. \ref{fig:p_distribution} shows the distribution of $p$ for our data.   Now, $p$ will range between 0 and 1, both inclusive. The reason can be very easily seen from the following:
\begin{itemize}
	\item $p=0$:  when both ego and alter converse in one particular language exclusively. 
	\item 	$p=1$: when either one uses one language and the other uses the other language exclusively.
	
\end{itemize}

We define an ego-alter pair to \emph{predominantly} converse in one language if $p<0.5$ . We can see from Fig. \ref{fig:p_distribution} that this is true for most of the ego-alters pairs. Moreover, the mean $p$ in our sample comes to be $0.31$. Indicating that most words exchanged between an ego-alter pair tend to be in one language or the other. We then decided to test the statistical significance of this finding with the following null and alternative hypothesis: 
\begin{equation}
H_0: p\geq0.5\\
\end{equation}
\begin{equation}
H_1: p<0.5
\end{equation}
We applied the \textit{t test} with $\alpha=0.05$ and obtained a p-value of $2.2\times 10^{-16}$. Hence, we reject the null hypothesis as the observed results are highly unlikely if $H_0$ is true. Thus,  we have indications that language reciprocity  exists in our dataset.

\subsection{Textisms} \label{sec:textism}
Darkin et al. define \emph{textism} as the different abbreviations, acronyms, slang, and emoticons typically used in text messages \cite{durkin2011txt}. We found two types of textisms in our text message dataset that we will discuss below.

 \textbf{Numeric Homophones: } Numeric Homophones are digits that are used in place of a word due to their acoustic similarity. Our survey showed that  $14.4\% \pm 6.58\%$  participants  regularly use numeric digits.  We note that the same practice was observed by Verheijen et. al. in their study of \texttt{WhatsApp} and \texttt{Facebook} chats of Dutch teenagers \cite{verheijen2016}. Some examples include using \textit{4} to replace \textit{for}  and using the digit \textit{7} in place of \textit{saath} (which means both the number 'seven' and `together' in Urdu). 
	
\textbf{Character Repetition: } We noticed that a large number of words have repetition of characters, e.g., \textit{pleasseeeeeeeee} instead of \textit{please} or \textit{yessss} instead of \textit{yes}. We wanted to find user's intention behind repeating characters. For this purpose, we conducted a small independent user study and asked our participants whether they followed this practice in their text messaging or not? And if they do, then what was the reason behind it? We got responses from $67$ participants ($33$ males and $34$ females). Our results showed that $67\%$ participants regularly follow this practice and the thematic analysis of their responses showed that $53.33\%$ do it to put emphasis on that word.

\begin{figure}[!b]
	
	\centering 
	\includegraphics[width=\linewidth]{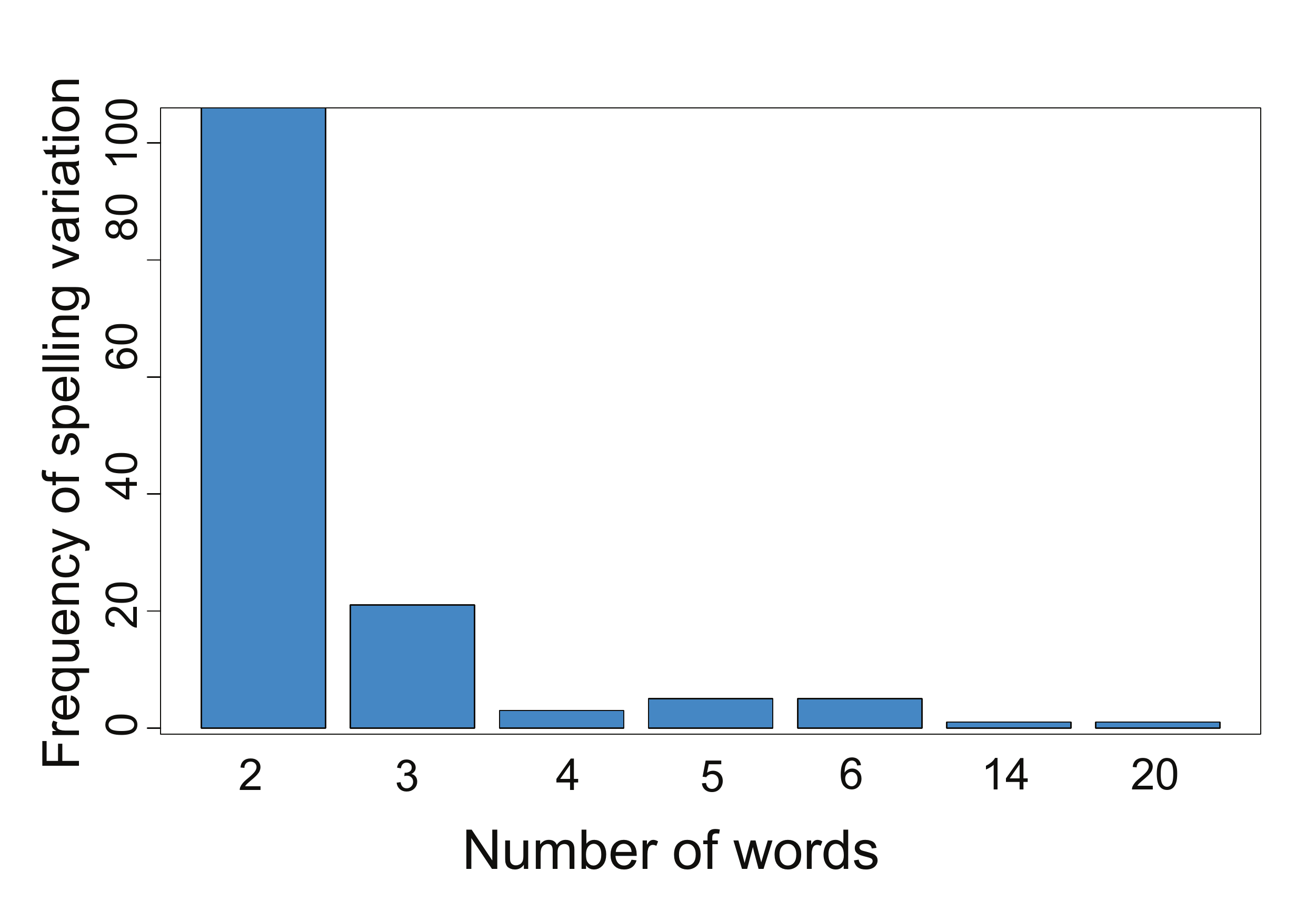}
	\caption{Plot showing frequencies of spelling variations. We can see that most words have $2$ spelling variations, while some words have as high as $20$ spelling variations.}
	\label{fig:spellingvariations}
\end{figure}

\section{Spelling Variations}\label{sec:spelling_variations}

Many Roman Urdu words have multiple spellings because it has evolved naturally and even the same person may write the same word with slightly different spellings at different times.  We wanted to better understand  these  \emph{spelling variations} in Roman Urdu. For this purpose, we first hand-labeled words having multiple variations in each user's profile separately, we  then combined files of all our participants, and finally  we enlisted  all the words with  multiple spelling along with their count. We found $586$ unique words with multiple spellings. Refer to  Fig. \ref{fig:spellingvariations} for further details. 

In a natural language processing (NLP) task on a writing scheme with non standard spelling rules, like Roman Urdu, a necessary preprocessing step would be to determine whether two strings $s_1$ and $s_2$ correspond to the same word or not.  An obvious first approach to solve the above mentioned preprocessing step is to apply a machine learning algorithm.  However, after eye balling the data, we noticed that most of these word-pairs vary in limited ways, for example users seemed to use `i' interchangeably with `e'. In order to systematically analyse their spelling variations, we designed an algorithm by modifying the \emph{Levenshtein Edit Distance} algorithm  \cite{levenshtein1966binary}.  Our algorithm enlists all spelling changes between two variations of the same word, more specifically it categorizes the change in spellings as either an  addition/deletion of a character; or changing one character to another. Our data analysis revealed the following:

\begin{description}

	\item \textbf{Add/Delete:}   We found that $61.65\%$ of the change operations were Add/Delete operations. Moreover, in about $90\%$ of the cases either a vowel or the character h or n was added or deleted as their sounds are close to certain Urdu alphabets. 
	
	\item \textbf{Replace:}  The most common replace operation was interchanging e and i in a word-pair which accounted for $25\%$ of such spelling change operations. This was again not surprising because the Urdu alphabet Chot\={i} ye is phonetically close to both 'i' and 'e'.
\end{description}

Hence, we showed that one can determine with reasonable accuracy whether two strings correspond to the same word or not, given the probability distribution of the various subtypes of variations. Such an algorithm can have many applications, such as Roman Urdu chatbots.
We next look at the two case studies which explore the applications of our dataset.

\section{Case Study 1: Ease in Text Entry} \label{sec:word_completion}

These days a wide variety  of applications are available for smartphones users, but  text entry is still the most common activity  on smartphones \cite{do2011smartphone, falaki2010diversity}. Hence, it is reasonable to assume that if we improve the usability of text entry on a smartphone, then we will also have significant impact on the overall usability of smartphone usage.  This is probably the reason why software developers as well as research community is continuously trying to develop improved virtual keyboards for smartphones. Typing on a virtual keyboard is more difficult because of their smaller size and because they lack tactile feedback  \cite{hoggan2008investigating}. Different techniques have been used to improve typing experience on virtual keyboards, however many existing virtual keyboards use language model to speed up typing by predicting the complete word based on the previously typed letters \cite{goodman2002language}.  There has also been extensive research on improving text entry of virtual keyboards, for example improving text entry  when user is stationary \cite{henze2012observational,rudchenko2011text, sears1993investigating,gunawardana2010usability}  and when user is  walking \cite{schildbach2010investigating, mizobuchi2005mobile}. However, to the best of our knowledge there has been no work that studies improving text entry for Pakistani users in the context of Roman Urdu.

\begin{figure} [t]
	\centering 
	\includegraphics[width=\linewidth]{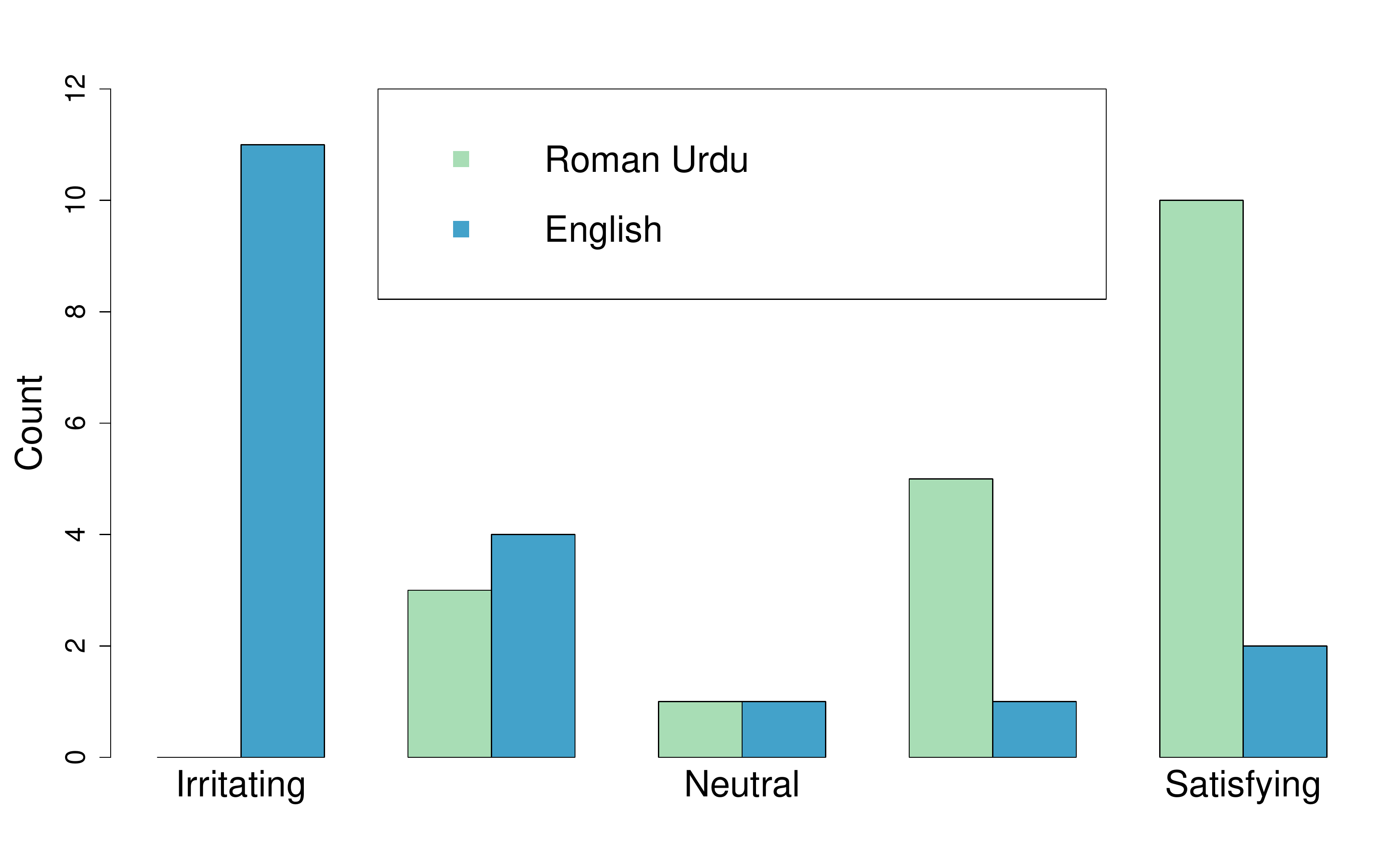}
	\caption{ Subjective opinion of $17$ participants about English word completion versus Roman Urdu word completion. We can see that users find completing words in a phone with English dictionary \emph{irritating}.}
	\label{fig:word_completion_experience}
\end{figure}
In order to assess  text entry needs of Pakistani users,  we conducted a pretest, in the form of a survey. Survey data was collected from 106 individuals, $43.4 \%$ of them being males while the rest were females. The age of these individuals ranged from $17$ to $45$ years with a mean of $24.5$ years.  Through the survey, we found that $71\%$ of our respondents type messages in Roman Urdu and $60\%$ feel that a specialized keyboard for Roman Urdu would be beneficial. One way to improve typing in Roman Urdu would be to improve auto-complete features on current keyboards as  they do not provide word completion on non standard languages like Roman Urdu.  This  argument was supported by the results of a survey of $17$ individuals which is summarized in  Fig. \ref{fig:word_completion_experience}.

We have observed that many Pakistani users,over time, manually enter Roman Urdu words in the phone dictionary to make their text entry easier. As most Pakistani people generally type in Roman Urdu, it intrigued us to estimate how much time will be saved  if the word completion algorithms are pre-trained with Roman Urdu dictionary. In this regard, we conducted two tests. First, we estimated how many words are completed by pre-training a word completion algorithm in a computer simulation. We then conducted a controlled experiment to measure the difference in time and subjective opinion when a dictionary is trained in Roman Urdu versus a standard smartphone dictionary available to Pakistani users. We will discuss them one by one below.

\subsection{Computer Simulation}
One of the most primitive ways to complete the spelling of a word, given a prefix, is a Radix Tree.  We used radix tree to estimate the increase in usability if the dictionary of smartphones uses dataset of local words. We conducted our experiment on three datasets, two of them were English datasets i.e. SMS corpus collected by Tagg \cite{tagg2013language} and $3000$ most commonly used words  in English given by Education First \footnote{\url{http://www.ef.com}}, and the third was the Roman Urdu dataset that we collected. We extracted unique words of all users from our Roman Urdu dataset. We divided our unique words data into two segments; $80\%$ data was used as training data while $20\%$ unique words used as our test data. We then built a radix tree using all words in the training set and then iteratively checked if each word in the training set is correctly completed through the same radix tree. As expected, radix tree built with Roman Urdu dataset outperformed in word completion by accurately completing $89\%$ words of the test data, while the accuracy of word completion when  the radix tree is built with with Tagg's \cite{tagg2013language} dataset and the top $3000$ words was significantly lower. Results of this experiment are given in Table \ref{tab:word_completion}.

\begin{table*} 
	\centering 
	\caption{Results of Word Completion Experiment using radix tree}
	
	\begin{tabular}{l|l|l}
		\hline
		\textbf{Dataset}   & \textbf{Words completed}    & \textbf{Words not completed}   \\ \hline\hline
		Roman Urdu Corpus & $19,370$  ($89\%$) & $2,404$    \\ \hline
		Tagg's English Corpus & $6,497$  ($30\%$)& $15,275$\\ \hline
		Top $3000$ most used English words  &$6,032$  ($28\%$) &$15,741$\\ \hline
		
	\end{tabular}
	\label{tab:word_completion}
\end{table*}

\subsection{User Study}

\begin{figure} [t]
	\centering 
	\includegraphics[width=\linewidth]{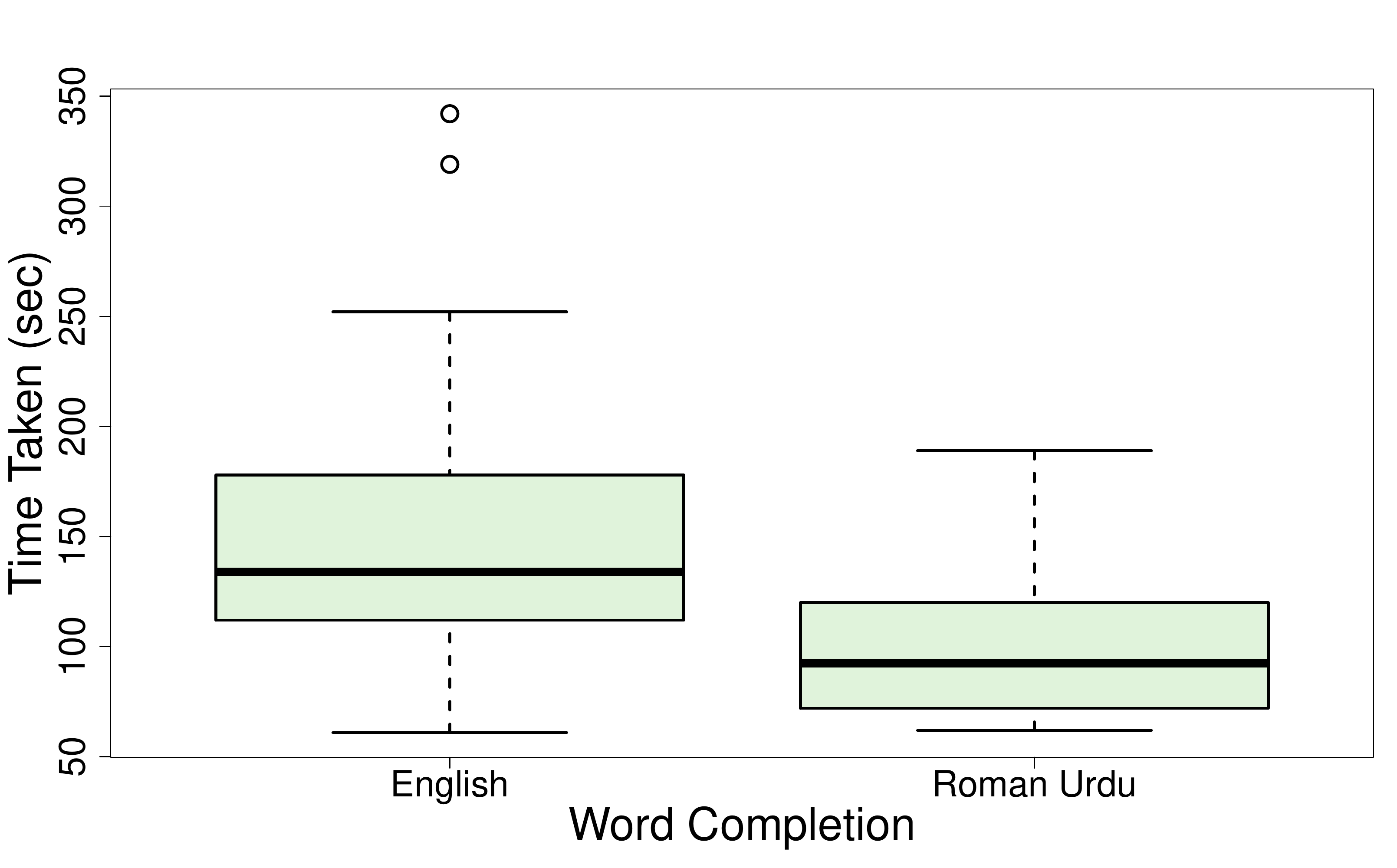}
	\caption{Time taken by  participants in  English word completion versus  Roman Urdu word completion. We can see  that the participants using Roman
Urdu word completion in general took less time to perform
this activity.}
	\label{fig:timeComparison}
\end{figure}

We also conducted a controlled experiment to measure the time taken when a user types on a smartphone with standard word completion versus a smartphone with Roman Urdu word completion. We designed our experiment as a \emph{between-group} experiment. We first selected $30$ participants ($15$ male and $15$  females) with ages between $24$ and $30$ years. We then randomly divided these participants into two groups of $15$ each.  Each group was asked to enter a piece of Roman Urdu text in a smartphone provided by the first author of this paper. This smartphone had a screen size of $5$ inches and was running \texttt{Android} operating system. One group entered the given text in a mobile phone with previously added Roman Urdu. While the second group entered the text with a default word completion (i.e. English). 

For both groups, we noted the time taken in seconds. The boxplot in Fig. \ref{fig:timeComparison} shows that the participants using Roman Urdu word completion in general took less time to perform this activity. We denote the mean time taken for English word completion as $\mu_e$ and we calculated it to be $161.92$ (SD= $84.70$ \footnote{Note: SD here refers to standard deviation}). Similarly, we denote the mean time taken for Urdu word completion as $\mu_u$ and we calculated it to be $100.43$ (SD= $33.34$). We next wanted to test if this difference is statistically significant by applying t-test with the following set of hypotheses:

\begin{center}

	\textbf{$H_0$: } $\mu_e=\mu_u$
	
	\textbf{$H_1$: } $\mu_e>\mu_u$
	
\end{center}   

The p-value was computed to be $0.01$ and we concluded that it is highly unlikely that we get this data if $\mu_e$ and $\mu_u$ are equal. 

Hence, a straightforward and simple way to improve the typing speed of an emergent users would be to use a text entry dataset  to train the word completion software. The only condition being that such users write their local language in Roman script or use a local variation of English.

\section{Case Study 2: Sociological Analysis} \label{sec:intimate}
In this case study we analyse the sociological aspects of text messaging. 
We started by looking at the top $20$ word bigrams that our participants have used in the text messages. It was surprising to notice a large proportion of intimate words. We divided the words into two categories: ordinary words and romantic/intimate words. For the top $20$ unique word bigrams, we found that $50\%$ words were of ordinary nature,whereas, $45.5\%$ were of intimate nature. The remaining $4.5\%$  words seemed to be part of marketing and promotional messages.  Encouraged by the high percentage of intimate bigrams we extracted frequently occurring words which depict intimacy.

\begin{figure} [!t] 
	
	\centering 
	\includegraphics[width=\linewidth]{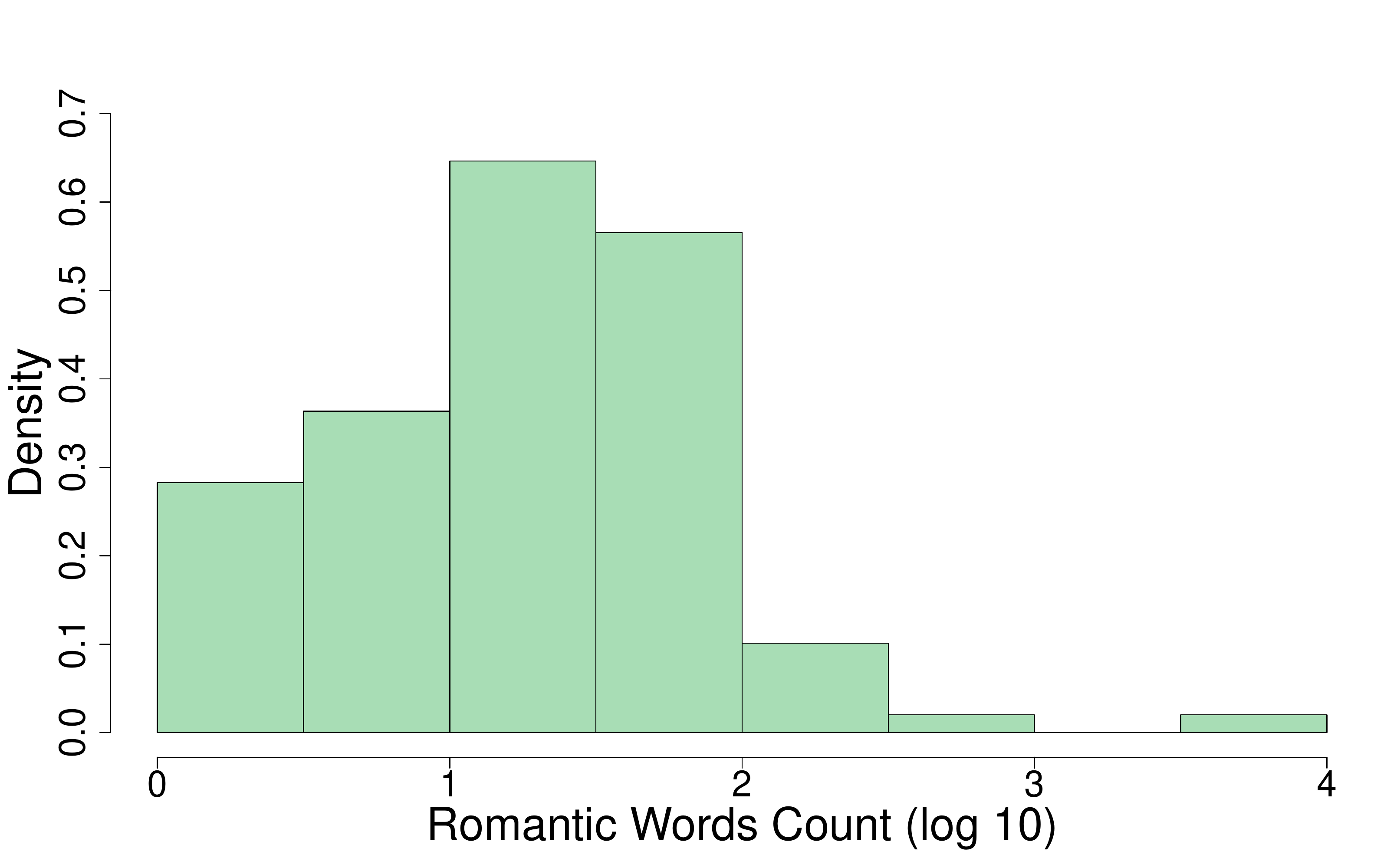}
	\caption{Distribution of romantic words used by our participants. We can see that most of our users used 100 or less words, while there are a few outliers with a high count.}
	\label{fig:filteredromanticboxplot}
\end{figure}

We counted such words for each individual. We found that there were $99$ participants who used intimate words.  The  number of such words ranged from $1-3559$ words in the files of those $99$ participants. The total number of words expressing intimate relations were $7129$. Fig.  \ref{fig:filteredromanticboxplot} shows the distribution of romantic words count per user. We recall from Section \ref{sec:data_collection} that many  participants deleted complete conversations with $1$ or $2$ of the alters, some of which were their intimate alters. Hence the difference in results of intimate and non-intimate alters as discussed in this section are likely to be more extreme in reality.
\begin{figure}[b!]
	
	\centering 
	\includegraphics[width=\linewidth]{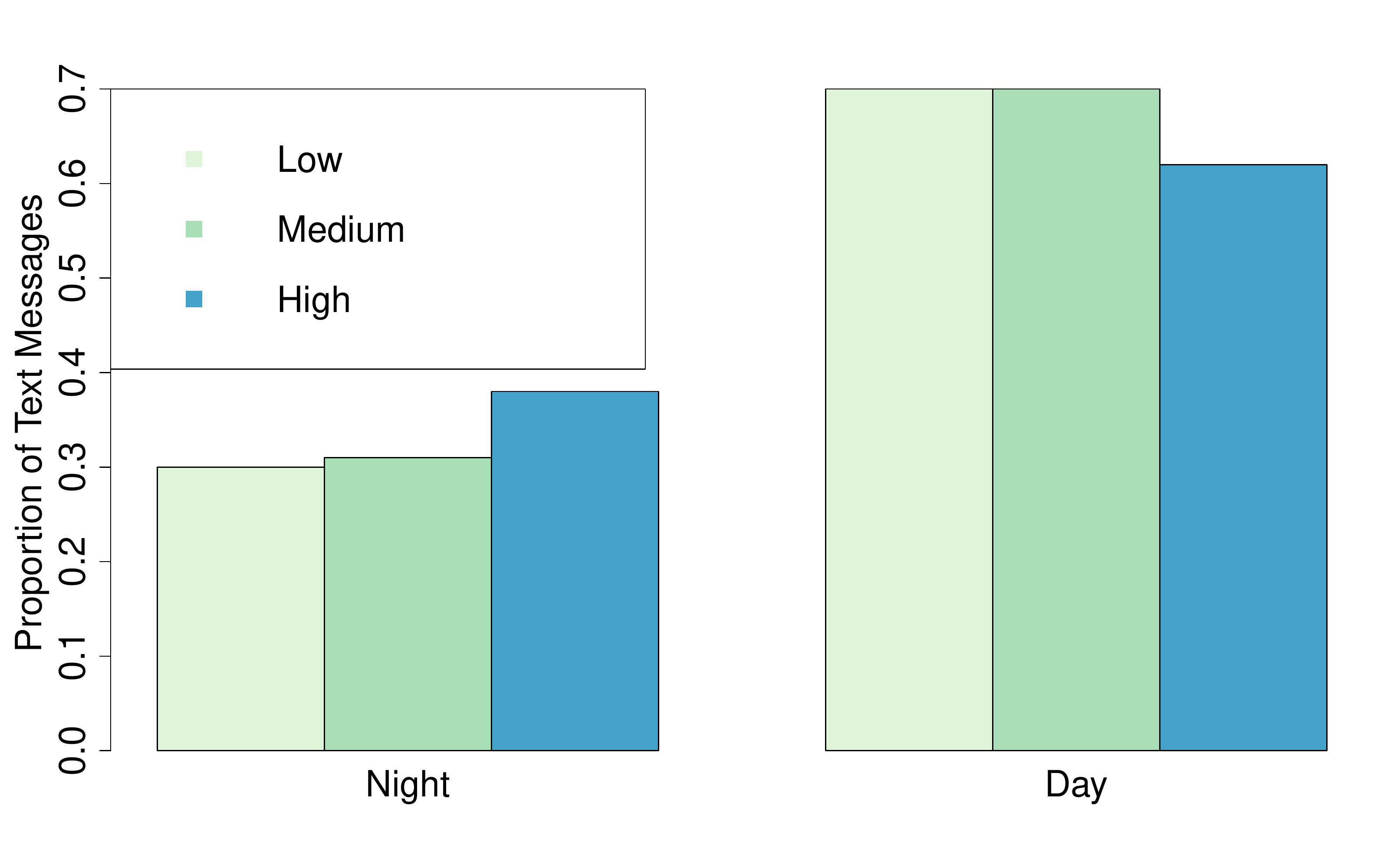}
	\caption{Comparison of the three groups of participants. We can see that the user with high number of romantic words communicate more than the other three groups at night. Night time is defined to start at 8:00 PM and ends at 7:00 AM}
	\label{fig:3groups}
\end{figure}

We grouped our participants into three disjoint sets based on the number of romantic words in their text messages. The three groups and the grouping criteria is given below:
\begin{enumerate}
\item \emph{Low romantic} group had $20$ or less romantic words.
\item \emph{Medium romantic} group had between $21$ and $80$ romantic words.
\item \emph{High romantic} had more than $81$ romantic words.
\end{enumerate}

We next checked the temporal characteristics of these text messages. Fig. \ref{fig:3groups} shows the average proportion of text messages sent or received for each user group. It shows that with increasing number of romantic words, users seem to communicate later in the evening.  In the Pakistani society generally it is easier for people to communicate with their romantic partners  late s

in evenings since that time is relatively more private \cite{nasim2016investigating}. A few studies investigate how romantic couples use smartphones \cite{jacobs2016} and choose communication media \cite{cramer2015}, albeit to the best of our knowledge such studies have not been conducted in this particular sociological setting. 
\begin{figure}[t]
	
	\includegraphics[width=\linewidth]{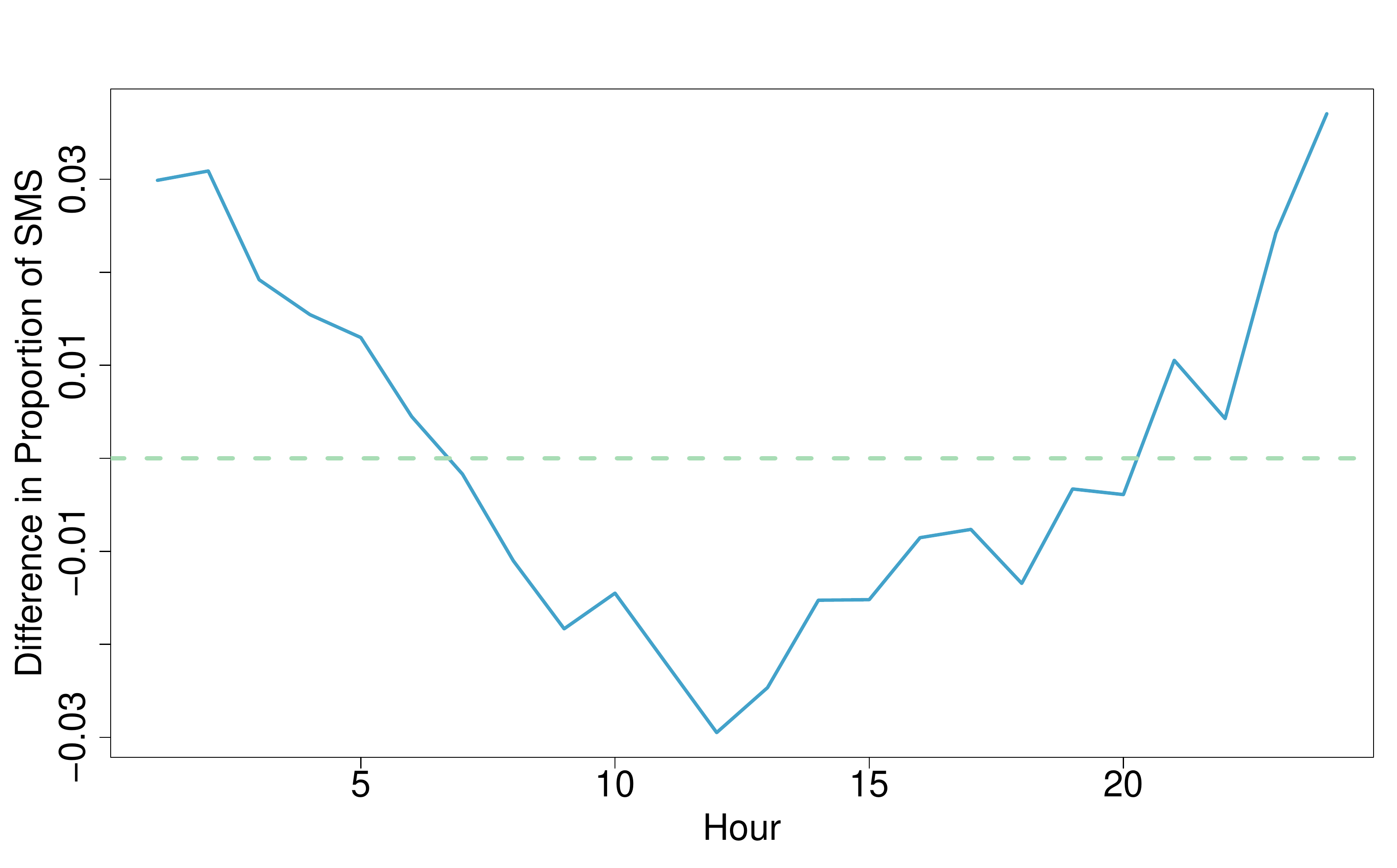}
	\caption{Time based comparison of the difference between the proportion of messages sent to intimate alters and   non-intimate alters. We can see that overall our users communicated more with the intimate alters after 8:00 PM. Note that the dashed horizontal line shows a difference of zero while positive values indicate that more messages were sent to intimate alters.}
	\label{fig:intimatealters}
\end{figure}


In order to ensure that the trend shown in Fig. \ref{fig:3groups} was not due to factors such as different chronotypes of the different groups of participants, we decided to select a set of ego-alter pairs, where the ego had sent five or more intimate words.  We restricted this to \emph{sent messages} in order to ensure that an unsolicited text message is not analysed. There were twenty six such participants (egos). We divided the contacts (alters) of this set into two disjoint subsets: intimate alters and non-intimate alters.  Here, \emph{intimate-alters} are the contacts with whom five or more intimate words were communicated and  \emph{non-intimate alters} are the remaining contacts. We calculated the proportion $p_i$ of messages that were exchanged with  intimate alters and the proportion $p_n$ that were exchanged with non-intimate alters. These proportion were calculated for each hour of the day. Next we computed their difference as $d=p_i-p_n$. We can see from Fig. \ref{fig:intimatealters} that  the same group of people tend to send more SMS to their intimate alters as compared to their non-intimate alters after 8:00 PM. To test this hypothesis, we applied t-test and found that the probability of an ego communicating with an intimate alter after 8:00 PM is more than his/her non intimate alters ( p-value $=0.003$). 

These results are supported by sociological theories that suggest that a modern man living in urban society is a part of several social groups known as social circles and he communicates with people in each social circle at a different pace during different times of the day \cite{simmel2012grossstadte}.

\section{Conclusions and Future Work}\label{sec:conclusions}

In summary, we looked at the structure, function, and possible benefits of SMS data of emergent users in Pakistan. We recall that we collected various derived data from SMS data of $116$  Pakistani students.    Moreover, no personal information such as names or phone numbers were collected. We also collected qualitative information from the same users through a survey. 

We found that most users use Roman Urdu to communicate on text messages. Two thirds of our participants think it is because they are able to express themselves better this way.  Moreover, the choice of language was not always consistent and we found to be dependent on the recipient of the message. Users also seem to adjust their language selection to match the language of the alter. Specifically, we found that some ego-alter pairs mostly communicate in English, while others mostly communicate in Roman Urdu.

Our work has three main conclusions: 
\begin{enumerate}
	\item  Text entry is one of the most common tasks performed on smartphones. When we conducted a survey about the text entry needs of Pakistani users, we were surprised to find that they did not feel the need for a specialized Urdu keyboard in its Arabic script, instead they felt the need for a Roman Urdu keyboard that helps them type faster on smartphones. This seems to indicate that a quick and easy way to improve text entry for emergent users is by training the word completion module using a dataset of local users' text messages. We showed that in this way the proportion of words completed  is higher than default word completion. Moreover, the satisfaction level of the users is also better when this is done.
	
	\item  We note here that the dataset we collected can help in creating Roman Urdu chatbots.  We recall that a chatbot is a software system that conducts conversations with human via textual or auditory methods.  Although, the first chatbot ELIZA was created about 50 years ago, it is only recently that they are changing human-computer interaction with Apple Siri, Google Now and Amazon Alexa. There has been a lot of research and development in chatbots and natural language processing. However, the spelling variation in Roman Urdu posed a fundamental problem for developing chatbots that can converse in Roman Urdu with the user. However, we have seen that it is not difficult to solve this preprocessing step. Since Roman Urdu has naturally evolved, there are no standardized spellings in place. However, in Section \ref{sec:spelling_variations}, we showed that the spelling variations follow certain patterns. Moreover, the probability of certain patterns occurring is significantly higher than others. Hence, it is not difficult to solve the problem of spelling variations. This information can be used in improving the design of many existing applications and can spin-off new applications, for instance Roman Urdu chatbots which can help emergent smartphone users achieve a number of tasks with greater ease. Moreover, a similar approach can be applied to other emergent users who use non standard spellings. We are in the process of designing a Roman Urdu chatbot which would help people to report domestic violence.
	
	\item  It was very surprising for us to discover many participants use SMS to communicate with their romantic partners.  In our initial study \cite{bilal2017roman}, about $10\%$ of our participants suggested that their relational messages consists mainly of  intimate messages.  We were intrigued as  romantic relationships are generally discouraged. We studied this more deeply by selecting words used in romantic conversations by hand. We discovered that many participants had high density of these words. We hence think that many young {men} and women have adopted text messaging for their more intimate conversation due to its cost efficiency and  privacy. It might be useful to study this large segment of users more deeply to better understand their needs and requirements. 
	
\end{enumerate}

In summary, this is the first such study of SMS data of a large but understudied population with significant benefits.  However, like any research study, there are certain aspects that threaten its validity. For example, the sample of participants in study is based on convenience sampling hence some results may not be generalized. Similarly, the results in Section \ref{sec:spelling_variations} are based on change operations detected by an algorithm, these operations might be different than how a human would classify  them, which also makes this a future research topic. Similarly, Section \ref{sec:intimate} might be underestimating the difference in communication patterns between intimate alters versus non-intimate alters as a number of our participants seemed to have deleted whole conversations taken place with a few of their alters as discussed in Section \ref{sec:data_collection}.

\section*{Acknowledgment}

We are thankful to the participants of this study which helped us in completing this work.

MN acknowledges support from ARC centre of Excellence for Mathematical and Statistical Frontiers, Australia.

\bibliographystyle{ieeetr}
\bibliography{myref}

\EOD

\end{document}